# Leveraging ERP Implementation to Create Intellectual Capital: the Role of Organizational Learning Capability

**Quang V. Nguyen**
School of Information Management
Victoria University of Wellington
Wellington, New Zealand
Email: quang.v.nguyen@vuw.ac.nz

**Mary Tate**
School of Information Management
Victoria University of Wellington
Wellington, New Zealand
Email: mary.tate@vuw.ac.nz

**Philip Calvert**
School of Information Management
Victoria University of Wellington
Wellington, New Zealand
Email: philip.calvert@vuw.ac.nz

**Benoit Aubert**
School of Information Management
Victoria University of Wellington
Wellington, New Zealand
Email: benoit.aubert@vuw.ac.nz

## Abstract

The extent to which enterprise resource planning (ERP) systems deliver value for organizations has been debated. In this study, we argue that the presence of appropriate organizational resources is essential for capturing the potential of ERP implementation. We investigate the relationship between ERP implementation and two organizational resources, specifically, Intellectual Capital (IC) and Organizational Learning Capability (OLC) to enrich the understanding of the way the value of ERP implementations can be realized. A sample of 226 manufacturing firms in Vietnam was surveyed to test the theoretical model. Structural equation modelling with partial least square method and two approaches for moderation analysis were used to analyze the data. The results indicate that ERP implementation scope has a positive impact on intellectual capital (IC). However, firms need to build a certain level of OLC to utilize ERP implementation for the enhancement of IC.

**Keywords**

IT business value, ERP implementation scope, Intellectual capital, Learning capability, Moderating effect.

## 1   Introduction

The benefits and the impacts of ERP systems for firms have long been the focus of much previous research (Schlichter and Kraemmergaard 2010). Some have argued that IT resources in general (Carr 2005), or ERP in particular (Seddon 2005) are commodities that are available to all firms and therefore unlikely to create competitive advantage.  Another possible explanation is it depends on how these resources are used or employed in organization (Legare 2002; Liang et al. 2010). Similarly, as proposed by Melville et al. (2004), the interrelationships between IT resources (e.g. ERP systems) and other organizational resources determine organizational performance, rather than the IT implementation alone.

Numerous studies have focused on the benefits of ERP (e.g. Chang et al. 2011; Ifinedo 2006; Shang and Seddon 2002) but have largely ignored the relationship between ERP and other organizational resources. These studies have attempted to categorize the benefits of ERP using various approaches, for example balanced scorecard (BSC) (Chang et al. 2011); analysis using five benefit categories: operational, managerial, strategic, IT infrastructure, and organizational (Shang and Seddon 2002); or three levels of benefits: individual, workgroup and organizational (Ifinedo 2006).  However, none of these frameworks consider the interaction of ERP implementation with other organizational resources to create benefits. This study views the impacts of ERP in a different way using IT business value framework of Melville et al. (2004). It focuses on the relationship between ERP and two other organizational resources: organizational learning capability (OLC) and intellectual capital (IC). The study attempts to answer two research questions: To what extent does the scope of ERP implementation lead to the enhancement of IC? What is the moderating effect of OLC on the relationship between the scope of ERP implementation and the enhancement of intellectual capital? The research questions are based on two notions. First, ERP implementation is considered as an opportunity for organizational learning and through which the sum of knowledge or intellectual capital is accumulated. Second, the examination of the relationship between ERP and other organizational resources (e.g. OLC and IC) is necessary for future studies on the possibility that ERP implementation leads to competitive advantage.

OLC is a concept involves the conditions and factors that facilitate an organization to learn (Dibella and Nevis 1998; Jerez-Gómez et al. 2005). Learning capability is an intangible resource that has an essential role in the enhancement of firm's performance and competitive advantage (Lei et al. 1996) in general and in the success of IT implementation (Ke and Wei 2006; Robey et al. 2002) in particular. Besides, organizations are considered as knowledge-based entities which utilize their knowledge to create value (Grant 1996) and to accumulate further knowledge (Vera and Crossan 2003). IC is the sum of knowledge of an organization that can be used to leverage the competitive performance of the firm (Youndt et al. 2004). Prior literature has posited that IT investment is associated with increases in organizational capital and IC (Brynjolfsson et al. 2002; Youndt et al. 2004). However, this general insight has not been tested with regard to understanding the benefits of ERP implementation.

The paper uses survey data to investigate these research questions. Results show that the three dimensions of ERP implementation scope (magnitude, breadth, and depth) have a positive impact on intellectual capital. The results also indicate that OLC moderates the relationship between ERP implementation scope and intellectual capital. The next sections discuss the literature, hypotheses development, method, results and discussion of the findings.

## 2　Literature Background

### 2.1　Resource based view and IT business value

The Resource Based View of the firm (RBV) (Barney 1991) indicates that firm performance is determined by its strategic resources. Strategic resources have the characteristics of rareness, value, imperfect imitability, and non-substitutability that support firms hold competitive position. Accordingly, because IT products are argued to be a commodity (Carr 2005), in order for IT resources to become valuable firm-specific assets, they need to be combined with other organizational resources and/or capabilities to lead to positive impacts and better organizational performance (Melville et al. 2004). The combination of IT resources and other organizational resources can be in the forms of complement and moderation (Wade and Hulland 2004). Resources are complementary when their combination leads to higher impact as compared with if using them separately (Teece 1986). Moderation occurs when some organizational factors influence the relationship between IT resources and organizational business value (Wade and Hulland 2004).

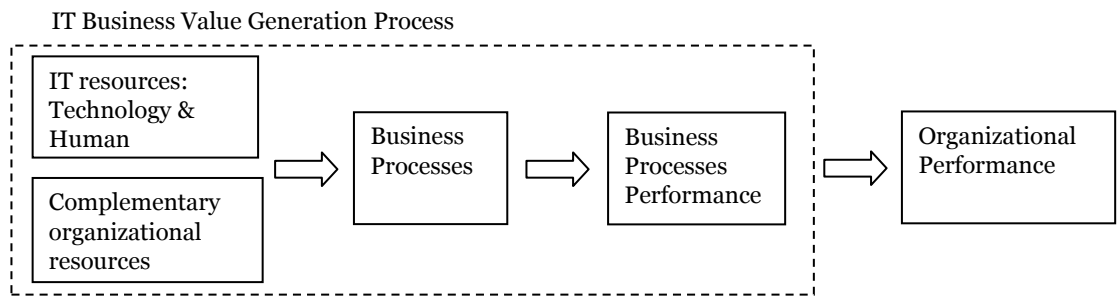

*Figure 1: IT business value generation process, extracted from Melville et al. (2004)*

In order to understand how IT contributes to organizational performance, Melville et al. (2004) conducted a substantial review of the prior research and proposed a model based on the RBV in order to explain the mechanism through which IT can bring value to organizations. As illustrated in Figure 1, both IT resources and other organizational resources are in the process of value creation for firms.

However, this is a high level model that does not consider the nature of the relationship between IT and specific organizational resources, nor does it evaluate the model with respect to particular examples of IT implementation such as ERP systems.

## 2.2 Two strategic resources of a firm: organizational learning capability and intellectual capital

### 2.2.1 Organizational learning capability

DiBella and Nevis (1998) propose that organizational learning capability (OLC) can be understood through the examination of two parts related to that capability. One part is about the learning orientations within an organization. Specifically, the learning orientations refer to how and what an organization learns. The other consists of facilitating factors that promote this organizational learning. Combining the two parts might assist our understanding of organizational learning as the whole. However learning orientations do not provide criteria for *evaluating* an organization's learning capability, they only help us *understand* and *describe* the learning processes that currently take place in an organization (DiBella and Nevis 1998). In other words, learning orientations only reveal the learning process in a descriptive way. By contrast, facilitating factors help to explain the learning in a normative manner that is about how an organization should learn. Put differently, if these factors are strong and widespread in an organization, the learning will occur easily and there will be more of it (DiBella and Nevis 1998). We posit that the ability of an organization to leverage the potential of an ERP implementation is determined by facilitating factors. In the same manner, Jerez-Gómez et al. (2005) say OLC is characterized by the conditions or facilitators for effective organizational learning.

OLC has been conceptualized as a multidimensional construct that comprises managerial commitment, systems perspective, openness and experimentation, and knowledge transfer and integration (Jerez-Gómez et al. 2005). Managerial commitment refers to the role of management in creating a culture of learning. Managerial people should hold a view that learning is of fundamental value. They should participate and also encourage employees to participate in learning. A system perspective denotes the ability to think broadly about the interdependency of organizational factors (Nevis et al. 1995). It is associated with creating a shared vision and mental models in an organization. Without any shared vision, individuals and departments are less likely to know what the organization's objectives are and how to measure and achieve them. For proactive learning, openness and experimentation are necessary for the organization to welcome new ideas (Senge 1990). Openness and experimentation are also associated with the notion of "unlearning". "Unlearning" is vital for organizational change, organizations need to "proactively question [their] long-held routines, assumptions, and beliefs" for the readiness of change (Sinkula et al. 1997, p. 309). Finally, knowledge transfer and integration represents the extent to which the organization has the ability to spread and integrate knowledge among their members (Jerez-Gómez et al. 2005). Knowledge sharing and integration is a key determinant for the establishment and capacity of organizational learning.

Learning capability is seen as a source of heterogeneity between organizations and formulates the competitive advantage through the accumulation of knowledge (March 1991). OLC is also a strategic resource because the competitive advantage of a firm in the future is determined by its ability to learn faster than its competitors (DeGeus 1988).

### 2.2.2 Intellectual capital

Intellectual capital (IC) is a broad concept that has a relation with intangible firm assets (Marr and Adams 2004). The term IC may be used with different meanings depending on the field or background that researchers are interested in (Marr and Chatzkel 2004). With a focus on the field of knowledge management rather than economics, we followed Youndt et al. (2004) who defined IC as "the sum of all knowledge an organization is able to leverage in the process of conducting business to gain competitive advantage" (Youndt et al. 2004, p. 337). Based on a literature review, Youndt et al. (2004) categorized IC into three dimensions: human capital, that refers to the knowledge, attitudes, skills, competences, commitment, and experience of organizational members; organizational capital, that involves institutionalized knowledge and codified experience residing in databases, manuals, structures, and processes; and social capital, that includes the knowledge embedded in the social relationships and networks among individuals, communities, or society.

Knowledge is seen as the most important strategic resource of the firm (De Carolis 2002). From the knowledge-based perspective, a firm itself is considered as an entity creating and applying knowledge (Grant 1996; Nonaka et al. 2000) and converting knowledge into competitive advantage (Kogut and Zander 1992). Knowledge resources enable a firm to gain competitive advantage because they have the characteristics of rareness, imperfect imitability, and non-substitutability (Barney 1991). In the knowledge based economy, the creation of value is no longer based on material and physical things but on information, knowledge, and brainpower (Stewart 1997). IC has been widely highlighted as an organizational resource and it is essential for the attainment of high organizational performance (Bontis 1999; Youndt et al. 2004).

## 2.3　ERP implementation scope and time length for impact

ERP systems are large, complex and frequently have "teething issues". Therefore the "scope" of ERP implementation and the length of time since its go-live are expected to be significant.

The scope of ERP implementation reflects the extent to which ERP systems are diffused within an organization and its business processes and has three dimensions: breadth, depth, and magnitude (Barki et al. 2005). The breadth of implementation indicates the extent to which the implementation of the system and business process reengineering (BPR) is diffused horizontally across the organization. The number of functional units, the number of sites (Nagpal et al. 2015) that are integrated by the system and BPR activities, among others, are examples of this dimension. For instance, Arvidsson et al. (2014) looked at integration between different functions and different mills (sites). Scope can also be expressed in terms of business processes (Fryling 2015), which usually translates into specific modules into the ERP implementation (Siswanto and Maulida 2014). Breath enables the integration of processes between different organizational functions (Brehm et al. 2001). The depth of implementation refers to the extent to which the implementation of the system and BPR is diffused vertically in the organization. This dimension can be measured by assessing the number of users of the system and the number of employees whose activities are changed due to BPR (Arif et al. 2010; Barki et al. 2005), or by the extent to which the system is used at different hierarchical levels of the organization (Hedman and Johansson 2009). Finally, the magnitude of ERP implementation represents how much the system changes employees' work and business processes. Magnitude was assessed by multiplying the percentage of activities reengineered with the extent of modification they were subjected to (Arif et al. 2010; Barki et al. 2005).

ERP systems are large, complex and typically include a wide range of modules. It is, therefore, important to note that the scope of ERP implementation relates to its subsequent impacts (Markus and Tanis 2000). The organizational impact of implementing a single module is likely to be lower than a more comprehensive implementation, regardless of how successfully the organization carries out its implementation.

An organization also needs a certain period of time to obtain the benefits of ERP implementation (Markus and Tanis 2000). After initial ERP implementation, this organization experiences the "second wave" of implementation (Deloitte Consulting 1998). This comprises several stages. The "stabilise" stage is the time when this organization has to deal with many issues to familiarize itself with the new system and business process changes. At the "synthesise" stage, the system comes back to a normal operation, and this organization looks for ways to improve its business processes by adding new functionality modules and motivating end-users to support the changes. Finally, the "synergise" stage is when this organization attains the optimization of business processes that is expected to lead to the enterprise transformation. On average, organizations need from nine to twelve months since the go-live of the system in order to achieve desired effects (Deloitte Consulting 1998).

## 3　Research Model and Hypotheses Development

Based on Melville et al.'s (2004) framework, and our literature on organizational learning capability, intellectual capital, and ERP implementation, we proposed a research model as shown in Figure 2.

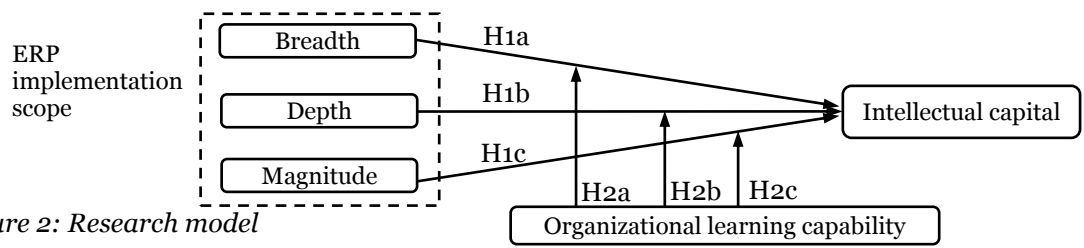

*Figure 2: Research model*

## 3.1　Relationship between ERP implementation and IC

ERP implementation can be seen as technology innovation so it is supposed to lead to change in an organization (Markus 2004). It involves not only technological but organizational and social aspects (Yeh and OuYang 2010). From this perspective, it is believed that the intellectual capital of a firm would be potentially influenced by ERP implementation in a number of ways.

First, ERP implementation may have a positive impact on human capital of a firm. During an ERP implementation, training programs are carried out for employees at all levels in the firm. Research on the impact of ERP systems has shown that these systems can result in improvements in some aspects of "human capital", e.g. enhancing the awareness and recall of job related information, and increasing the effectiveness in job and employee productivity (Gable et al. 2008). In an ERP implementation, employees might gain experience through their interaction with the ERP system (Clark et al. 2009). As a result, they can gain a better understanding of the ERP software and business processes.

Second, the organizational capital of a firm may be affected by an ERP implementation. According to Deloitte Consulting (1998), the main purposes of ERP implementations are: (1) to automate and integrate business processes, (2) to share common data and practices across the entire enterprise and (3) to produce and access information in a real-time environment. With ERP implementations, firms would like to document business processes and practices, standardize the processes according to those embedded in the software applications, convert the data into databases, and make all of these become daily activities and routines. These actions help embed the organizational knowledge and practices into firms routines. Similarly, as Shang and Seddon (2002) and Chang et al. (2011) remark, the benefits of ERP systems include cycle time reduction, data quality improvement, and better performance control. When these benefits become routine they are considered as organizational capital.

Third, an ERP implementation may also affect the social capital of a firm. The ERP implementation is associated with the mutual coordination and cooperation among individuals and workgroups (Ifinedo 2006). ERP systems improve communication and make business activities more integrated. As a result, the systems support users in working together, exchanging information, and creating and maintaining their relationships (Lengnick-Hall et al. 2004). The ERP implementation is expected to open the opportunity for a change in organizational culture and a common vision leading to the evolvement of individual relationships (Shang and Seddon 2002). The impact may also happen in the relationship between the firm and its external parties because ERP implementation is associated with the mutual understanding between ERP vendors, consultants and firms (Ifinedo 2006; Markus and Tanis 2000).

We therefore propose that the scope of ERP implementation has a positive effect on intellectual capital. However, as we discussed above, there are differences in the scope of ERP implementation which will have an impact on the amount of intellectual capital that is available to be captured by the organization. A broader scope, involving more modules, business processes, and business units, will create more opportunities for the firm to create intellectual capital, specifically:

H1a: the breadth of ERP implementation scope has a positive effect on intellectual capital

H1b: the depth of ERP implementation scope has a positive effect on intellectual capital

H1c: the magnitude of ERP implementation scope has a positive effect on intellectual capital

## 3.2　The moderating effect of organizational learning capability

It is commonly advised to consider organizational learning capability when studying the adoption of innovation technologies in an organization (Attewell 1992; Robey et al. 2002). Attewell (1992) described an IT adoption as a learning process. Implementation of a new information system (IS) like ERP is associated with learning that occurs throughout the whole organization. The ERP implementation involves a change in the organization and requires it to overcome knowledge barriers. When implementing an ERP system, the organization must learn to bridge the gap between what it has previously known and what the new IS requires it to know, in order to understand and use the new IS effectively (Attewell 1992), and to implement new processes with this new IS (Robey et al. 2002). Besides, an organization is considered as an entity creating and applying knowledge (Nonaka et al. 2000), therefore, if an ERP implementation is viewed as "a lesson" that the organization learns to accumulate its knowledge or intellectual capital, it means that organizational learning capability plays an essential role in regulating this learning.

Furthermore, the relationship between IT resources and business value creation can be understood by taking a look at contingent organizational factors (Wade and Hulland, 2004). According to the contingency theory, an organization can get a higher performance when there is a fit between organizational characteristics and contextual factors. In brief, "the contingency approach says that the effect of one variable on another depends upon some third variable" (Donaldson 2001, p. 5). In the IS domain, as Umanath (2003) remarks, fit as contingency can be examined in form of moderation analysis. A moderating effect indicates that the direction and/or the strength of the relationship between an independent variable and a dependent variable are influenced by a moderator (Baron and Kenny 1986). In the context of ERP implementations, as discussed above, an organization is a learning entity thus its learning capability likely moderate the effect of ERP implementation on intellectual capital. This point is supported through the fact that organizational learning capability facilitates the success of business system implementation (Ke and Wei 2006; Lee et al. 2007; Lin 2008). Therefore (H2): The relationship between the scope of ERP implementation and intellectual capital is moderated by organizational learning capability. Specifically:

H2a: The relationship between the breadth of ERP implementation and IC is moderated by OLC

H2b: The relationship between the depth of ERP implementation and IC is moderated by OLC

H2c: The relationship between the magnitude of ERP implementation and IC is moderated by OLC

## 4 Research method

The research model was tested using a set of empirical data collected through a survey conducted in Vietnam. This study employed the existing measures for the constructs used in the model and used Partial Least Square (PLS) modelling technique for data analysis.

### 4.1 Measurement

Two second-order constructs (intellectual capital and organizational learning capability) and three first-order constructs (breadth, depth, and magnitude of ERP implementation) were used in the study.

As for the measurement of IC, two widely-used measures for IC developed by two scholars: Bontis (1998) and Youndt et al. (2004) would be considered. Although the two measures focus on the same concept, they are distinct. While the measure developed by Bontis (1998) includes many items that are mainly associated with the outcome measurement, the measure proposed by Youndt et al. (2004) mainly reflects the current "state of being of IC" (Isaac et al. 2010). Because this study focuses on the effect of ERP implementation scope on the stock of knowledge of an organization, we adopted the scale items of Youndt et al. (2004). With the purpose of investigating the effect of ERP implementation on IC at broader level relationship, we specified IC as a second order construct consisting of three sub constructs: human capital (HC), organizational capital (OC), and social capital (SC). Although IC might be conceptualized as a reflective-reflective construct (Hsu and Sabherwal 2012), this study argues that IC is reflective-formative construct because its three dimensions are clearly distinct and independent: they describe unique aspects of the construct; they are not interchangeable; and if one dimension is dropped the conceptual domain of IC may be altered (Jarvis et al. 2003; MacKenzie et al. 2011).

Measurement of OLC was based on Jerez-Gómez et al. (2005) scale. OLC is a focal construct comprising four dimensions that demonstrate the attributes of a learning organization: managerial commitment (MC), system perspective (SP), open and experimentation (OP), and knowledge transfer and integration (KW). Although the authors conceptualized these dimensions as reflective indicators of the focal construct, this study considered the four dimensions as formative variables since they also are not interchangeable and might vary independently (Jarvis et al. 2003; MacKenzie et al. 2011).

The sub-dimensions of IC and OLC were measured on reflective perception-based items using seven-point Likert scales from 1 (strongly disagree) to 7 (strongly agree). There are fourteen items for IC and sixteen items for OLC. Additionally, one global item measuring the essence of each of these two constructs were included in order to assess their convergent validity (Hair et al. 2014).

Finally, ERP implementation scope was measured utilizing the scale proposed by Barki et al. (2005). Seven items were used to measure the breadth (BRE), depth (DEP), and magnitude (MAG) of ERP implementation scope. Most of these items use a ratio scale. The items of the research model constructs are presented in the Appendix 1.

### 4.2 Questionnaire construction

The questionnaire was designed with two parts: (A) the scope of ERP implementation, (B) the information of IC and OLC. The original questionnaire was in English. It was then translated into Vietnamese. A back-translation was used to ensure the equivalence of meanings between the two versions. A pre-test was done with five MBA students and two ERP implementation consultants in Ho Chi Minh City, Vietnam. As a result, modifications regarding wording, clarity, and sequence of questions were made. One more scale was added to the measure of ERP implementation breadth and two scales of the measure of ERP implementation depth were converted to the ratio of the original scale as compared with the number of employees of the firm.

### 4.3  The Sample

The sample frame included a list of 627 manufacturing firms that had implemented ERP systems in Ho Chi Minh City and Dong Nai province. These places are among five places of Vietnam which have the highest e-business index (VECOM 2013). The firm list was established from two sources: ERP providers and 2013 Vietnam Business Directory. Two respondents in each firm were required to answer the questionnaire: an IT manager (for part A) and another executive manager (for part B). The mail questionnaire survey was conducted over a period of eight weeks including two follow-up reminder telephone calls. A total of 242 questionnaires were returned. 226 usable questionnaires were retained for data analysis. Sixteen questionnaires were excluded because they had a significant number of incomplete items and/or the period that the ERP had been implemented was less than one year. Fifteen questionnaires, despite having missing values in descriptive part, were still retained because they had no effect on the regression results. The sample comprised 95 (42 percent) limited liability firms and 83 (36.7 percent) joint stock firms in other types of business (partnership, privately owned, and others). In terms of size, 54 firms (23.9 percent) had 500 employees or more, 61 (27 percent) had between 300 to 500 employees, 96 (42.5 percent) had between 100 and 300 employees, and 15 (6.6 percent) had fewer than 100 employees. Most of the firms in this sample are in the industry of electrical products (14.6 percent), food and beverage (13.7 percent), and construction materials (11.5 percent). While still not having transiting completely to a market economy, Vietnam has introduced significant market-oriented reforms (Athukorala 2006) and enjoyed significant benefits from a free-trade agreement with the U.S. (McCaig 2011). State-owned firms, which represented half of the country manufacturing output in 2000 only represented 18.5% of output ten years later. By 2009, 43% of manufacturing output came for foreign invested firms (Athukorala and Tien 2012). Results are therefore likely to be representative of manufacturing sectors in other countries with open economies.

### 4.4  Data analysis techniques

There are two different statistical approaches to estimate structural equation models: covariance-based SEM and variance-based partial least squares (PLS) modelling (Haenlein and Kaplan 2004).

In covariance-based SEM approach, an estimation function is used to minimize the difference between the sample covariance matrix and the hypothesized covariance matrix of the theoretical model (Chin and Newsted 1999; Hair et al. 2012). In contrast, PLS approach focuses on target dependent variables in the model and attempts to maximize the variance of the dependent variables that is explained by the independent variables (Haenlein and Kaplan 2004).

Although covariance-based approach is broadly used for the assessment of SEM, PLS is also an appropriate technique in doing SEM-based analysis under certain circumstances, such as the presence of formative measures, the complicatedness of research model, and the conditions of normal distribution (Chin and Newsted 1999). Based on the properties of our data and our research model, we view PLS approach as the most appropriate method for this study. The data set of this study had a number of items that were not normally distributed. The structural model is relatively complex with the presence of higher-order and formative constructs as well as moderating effects.

The model has two second-order constructs with formative dimensions and three moderating relationships. A redundancy analysis technique was used to assess convergent validity of two formative constructs (Hair et al. 2014). To evaluate the structural model and moderating effects, the repeated indicator approach and latent variables scores (Hair et al. 2014), and the approaches for moderation analysis suggested by Henseler and Fassott (2010) were used.

## 5  Results and Findings

### 5.1  Measurement models

The scales were assessed by confirmatory factor analysis (CFA) using SmartPLS (Ringle et al. 2005). Apart from two items (one of HC, one of SC) with loadings below 0.7 that were removed, all the remaining items had loadings on their designated constructs higher than 0.7 and greater than their loadings on any other constructs (Chin 1998). As shown in Table 1, the composite reliability (CR) scores are above 0.8, the square root of average variance extracted (AVE) of each construct is greater than its highest correlation with any other construct. The results suggest that all first order constructs have discriminant and convergent validity.

|     | AVE   | CR    | BRE   | DEP   | MAG   | HC    | OC    | SC    | MC    | SP    | OP    | KW    |
|-----|-------|-------|-------|-------|-------|-------|-------|-------|-------|-------|-------|-------|
| BRE | 0.795 | 0.886 | 0.891 |       |       |       |       |       |       |       |       |       |
| DEP | 0.820 | 0.901 | 0.246 | 0.906 |       |       |       |       |       |       |       |       |
| MAG | 0.757 | 0.904 | 0.126 | 0.364 | 0.870 |       |       |       |       |       |       |       |
| HC  | 0.653 | 0.882 | 0.279 | 0.325 | 0.348 | 0.808 |       |       |       |       |       |       |
| OC  | 0.671 | 0.890 | 0.478 | 0.437 | 0.307 | 0.466 | 0.819 |       |       |       |       |       |
| SC  | 0.678 | 0.894 | 0.380 | 0.404 | 0.278 | 0.506 | 0.558 | 0.823 |       |       |       |       |
| MC  | 0.671 | 0.911 | 0.480 | 0.392 | 0.333 | 0.448 | 0.514 | 0.376 | 0.819 |       |       |       |
| SP  | 0.715 | 0.882 | 0.419 | 0.321 | 0.228 | 0.363 | 0.508 | 0.435 | 0.545 | 0.845 |       |       |
| OP  | 0.714 | 0.909 | 0.434 | 0.367 | 0.331 | 0.394 | 0.519 | 0.334 | 0.558 | 0.446 | 0.845 |       |
| KW  | 0.712 | 0.908 | 0.385 | 0.400 | 0.256 | 0.548 | 0.535 | 0.386 | 0.571 | 0.489 | 0.496 | 0.844 |

*Note: Diagonal values are square root of construct's AVE.*

*Table 1. Average variance extracted (AVE), correlation, and composite reliability (CR) of constructs.*

Repeated indicator approach and a global item were used to evaluate convergent validity of two second order constructs IC and OLC. Using the redundancy analysis technique suggested by Hair et al. (2014), for each construct, all indicators of the first order constructs were assigned to the corresponding second order construct and a path between the second order construct and the global item was established. The structural path coefficient between OLC and its global item is 0.874, and is 0.824 for IC. The values of above 0.8 support the convergent validity of formative constructs (Chin 1998; Hair et al. 2014). Additionally, variance inflation factor (VIF) values of first order constructs of IC and OLC that vary from 1.439 to 1.915, all lower than the threshold value of 5 indicate the absence of multicollinearity among the formative dimensions (Hair et al. 2014).

## 5.2 Structural model

In order to assess the structural model, two-stage approach was used in combination with repeated indicators approach (Hair et al. 2014)1. In the first stage, two second order constructs OLC and IC were measured by the indicators of their first order constructs, then the PLS algorithm was applied to obtain the latent variables' scores (LVS). In the second stage, LVS of latent constructs were used to estimate the path coefficients in structural model and the moderating effects.

The structural model has three predictors (BRE, DEP, and MAG), one moderator (OLC), and one dependent variable (IC). We used both of two approaches suggested by Henseler and Fassott (2010) to analyse the model: group comparison and product term. In the group comparison approach, based on the LVS scores of OLC, two subsamples were extracted. Observations having OLC's LVS within upper third were specified as high OLC group; observations with OLC's LVS being within lower third belonged to low OLC group; the remaining observations were not assigned to any group. As a result of this division, the size of each sub sample was reduced to 75 observations. After that the same model was estimated for these two subsamples, path coefficients were compared.

As for the product term approach, two models were used: the first model had the direct effects of BRE, DEP, MAG, and OLC on IC and the second model additionally included three interaction terms (BRE*OLC, DEP*OLC, and MAG*OLC). The interaction terms were calculated as the products of the scores of OLC with the scores of BRE, DEP, and MAG.

For both of the two approaches, to determine path coefficients and their significance we used PLS algorithm and bootstrap technique (bootstrap samples of 5000, the size of each sample is equal the number of examined observations (Hair et al. 2014)).

As shown in Table 2, according to group comparison approach, the results from the whole sample (without the consideration of OLC) illustrate the positive impacts of all dimensions of ERP implementation on IC. Interestingly, there is no such relationship in low OLC group. Conversely, the group with high OLC shows that the breadth and magnitude of ERP implementation has a positive

---
[1] All indicators were standardized before estimating the structural model (Henseler and Fassott 2010).

impact on IC. Therefore, using this approach, all hypotheses (except H2b) are supported. The results from product term approach show that all dimensions of ERP implementation scope have a significant effect on IC. The significant effect of OLC on IC indicates a quasi moderation rather than pure moderation of OLC (Helm and Mark 2010). The presence of the moderation effects in Model 2 shows that only the magnitude of ERP implementation has a significant moderation effect with OLC on IC. Therefore, while the results support all main hypotheses (H1a, H1b, H1c), only one of three moderating hypotheses is supported (H2c). The moderating effect proves to be relatively small, with the size effect $f^2 = [0.537-0.517]/[1-0.537] = 0.0432$

| Path (group comparison) | Path coefficient | | | Path (product term) | Path coefficient | |
|---|---|---|---|---|---|---|
| | Basic model | Low OLC | High OLC | | Main effects | Interaction model |
| BRE → IC | 0.361** | -0.057 n/s | 0.200* | BRE → IC | 0.155 ** | 0.144 ** |
| DEP → IC | 0.307** | 0.018 n/s | 0.163 n/s | DEP → IC | 0.175 ** | 0.164 ** |
| MAG → IC | 0.220** | 0.091 n/s | 0.268* | MAG → IC | 0.124 * | 0.156 ** |
| R-square (IC) | 0.396 | 0.013 | 0.168 | OLC → IC | 0.466 ** | 0.492 ** |
| | | | | BRE*OLC → IC | | 0.043 n/s |
| | | | | DEP*OLC → IC | | 0.027 n/s |
| | | | | MAG*OLC → IC | | 0.117 * |
| | | | | R-square (IC) | 0.517 | 0.537 |

*Note: * Significant at 0.05   ** Significant at 0.01   n/s not significant*

*Table 2. Structural model results*

# 6 Discussion

Overall, the findings above show that three dimensions of ERP implementation scope have a positive impact on intellectual capital. The results also indicate that OLC more or less moderates the relationship between ERP implementation scope and intellectual capital. As a result of the group comparison approach for moderation analysis, we can see that firms with low OLC level are likely to have no effect of ERP implementation on IC (regardless of the scope of the implementation). However, in high OLC group the breadth and magnitude of ERP implementation (which create more opportunities for creating knowledge) have a positive effect on IC.

When we consider the analysis using the product term approach, only *magnitude* of ERP implementation shows an interaction effect with OLC on IC. The *breadth* and *depth* of ERP implementation reflects the horizontal and vertical diffusion of technology and BRP across the firm. These dimensions appear to have minimal interaction with OLC. The *magnitude* of ERP implementation represents the extent to which the system changes employees' work and business process automation (Barki et al. 2005). We suggest that magnitude of ERP implementation may have been significant in our study because this is the dimension most associated with organizational change and transformation. A higher proportion of modified activities of employees, and more automated business processes (i.e., lack of customization) thanks to an ERP implementation imply that the firm has the learning capability to deal with and embrace change in order to use the capabilities of the ERP system they have implemented.

At a broad theoretical level, this study based on IT business value creation using the lens of RBV to examine the impact of ERP implementation on intellectual capital as well as the moderating role of organizational learning capability. The findings of this study actually enrich our understanding of the relationship between IT resource (i.e., ERP systems) and other organizational resources. Practically, this study provides evidence that ERP implementation can create strategically valuable intellectual capital, provided the organization has the OLC necessary to capture the potential knowledge creation of the ERP system. Therefore, firms need to build a substantial level of OLC to fully utilize the value of ERP systems and need to pay attention particularly to the *magnitude* dimension of ERP implementation in the enhancement of intellectual capital.

---

[2] $f^2 = [R^2(\text{interaction model}) - R^2(\text{main effects model})]/[1 - R^2(\text{interaction model})]$; Interaction effect sizes are small if 0.02, medium if 0.15, and large if 0.35 (Cohen 1988).

## 7　Limitations and Further Research

This study only focuses on the relations between ERP implementation scope and the two organizational resources IC and OLC. As a consequence, a full list of possible ERP benefits that have been identified in prior studies (e.g. Chang et al. 2011; Ifinedo 2006; Shang and Seddon 2002) is not the emphasis in this study. It could be a path for future research. Additionally, with such a focus our study only reveals a part of the bigger picture of the mechanism through which ERP systems bring value to firms. Further research may include other organizational aspects into a more comprehensive model to explain how an ERP implementation leads to a superior organizational performance or competitive advantage.

This study used a set of data collected at one point of time. Although sampled firms were collected with the time length of at least one year of ERP operation, this may not reveal the complete phenomenon. A longitudinal study may be an alternative approach to have more insights into this phenomenon. We surveyed firms in Vietnam manufacturing industry, further research may evaluate the model with the data collected from other industries to make the results more generalizable. Additionally, the scope of ERP implementation used in this study was measured by the three generic dimensions, thus the research findings may not reflect the effectiveness of a particular ERP type. However this study can be replicated in other settings different from Vietnam manufacturing firms.

## 8　Conclusions

ERP systems represent enormous investments and disruptive changes for organizations, and many do not deliver all the expected benefits. This study confirms previous studies that show that technology alone cannot deliver strategically valuable resources for organizations. Nevertheless, organizations with the capability to learn and embrace change can leverage ERP implementations to create valuable intellectual capital that can contribute to improved competitiveness and organizational performance. This current study provides empirical analysis in support of the Melville et al. (2004) IT business value framework, which emphasizes the relationship between IT resource and other organizational resources. Practitioners' understanding of the impact of ERP implementations on organizational knowledge under organizational learning facilitators is further enhanced through this study.

## 9　References

# Appendix 1

Items in the intellectual capital scale

| | |
|---|---|
| Human capital | Employees of my organization are highly skilled<br>Employees of my organization are widely considered the best in our industry<br>Employees of my organization are creative and bright<br>Employees of my organization are experts in their particular jobs and functions<br>Employees of my organization develop new ideas and knowledge |
| Organizational capital | My organization uses patents and licenses as a way to store knowledge<br>Much of my organization's knowledge is contained in manuals, databases, etc.<br>My organization's culture (behaviours, stories, rituals) contains valuable ideas and ways of doing business, etc.<br>My organization embeds much of its knowledge and information in structure, systems, and processes |
| Social capital | Employees of my organization are skilled at collaborating with each other to diagnose and solve problems<br>Employees of my organization share information and learn from one another<br>Employees of my organization interact and exchange ideas with people from different areas of the organization<br>Employees of my organization partner with customers, suppliers, alliance partners, etc., to develop solutions<br>Employees of my organization apply knowledge from one area of the organization to problems and opportunities that arise in another. |
| Global item | Overall, my organization acquires the knowledge necessary for doing business that resided in people, mechanisms and structures, and relationships |

Items in the organizational learning capability scale

| | |
|---|---|
| Managerial commitment | The managers frequently involve their staff in important decision making processes<br>Employee learning is considered more of an expense than an investment.<br>The organization's management looks favourably on carrying out changes in any area to adapt to and/or keep ahead of new environmental situations.<br>Employee learning capability is considered a key factor in this organization<br>In this organization, innovative ideas that work are rewarded |
| Systems perspective | All employees have generalized knowledge regarding this organization's objectives<br>All parts that make up this organization (departments, sections, work teams, and individuals) are well aware of how they contribute to achieving the overall objectives<br>All parts that make up this organization are interconnected, working together in a coordinated fashion |
| Openness and experimentation | This organization promotes experimentation and innovation as a way of improving the work processes |
| | This organization follows up what other organizations in the sector are doing, adopting those practices and techniques it believes to be useful and interesting |
| | Experiences and ideas provided by external sources (advisors, customers, training firms, etc.) are considered a useful instrument for this organization's learning |
| | Part of this organization's culture is that employees can express their opinions and make suggestions regarding the procedures and methods in place for carrying out tasks |
| Knowledge transfer and integration | Errors and failures are always discussed and analyzed in this organization, on all levels |
| | Employees have the chance to talk among themselves about new ideas, programs, and activities that might be of use to the organization |
| | In this organization, teamwork is the usual way to work |
| | The organization has instruments that allow what has been learnt in past situations to remain valid, although the employees are no longer the same |
| Global item | Overall, my organization has adequate factors and conditions to facilitate learning |

Items in the ERP implementation scope

| | | |
|---|---|---|
| Breadth | ERP breadth [a] | 1= single site<br>2= multiple sites in one region<br>3= multiple sites in several regions<br>4= multiple sites in multiple regions across nation<br>5= multiple regions, international |
| | Business process reengineering (BPR) breadth | 1= small number of people within a dept.<br>2= a department<br>3= more than one department<br>4= a region<br>5= more than one region |
| Depth | ERP depth [b] | The percentage of users of the ERP software |
| | BPR depth [b] | The percentage of employees whose activities changed |
| Magnitude | Business process automation (BPA) | (% of processes that are automated after ERP) – (% of processes that were automated before ERP) |
| | BPR magnitude | (% of activities in reengineered processes that were modified) * (extent of modification of activities 1-10) |
| | ERP customization [c] | Extent of modification done to ERP to customize the software (from 1-10) |

a. One more scale was added to the ERP breadth item as a result of the pre-test
b. These scales were converted to the percentage as a result of the pre-test
c. Reverse item

# Copyright